# Incorporating Missing Data Considerations into Sample Size Calculations for Developing Clinical Prediction Models

Glen P. Martin[1]; Siân Bladon[1]; Rebecca Whittle[2,3]; Molly Wells[1]; Gary S. Collins[2,3]; Richard D. Riley[2,3]

1. Division of Informatics, Imaging and Data Science, Faculty of Biology, Medicine and Health, University of Manchester, Manchester, UK
2. Department of Applied Health Sciences, College of Medicine and Health, University of Birmingham, Birmingham, UK
3. National Institute for Health and Care Research (NIHR) Birmingham Biomedical Research Centre, UK.

**Running Title:** Incorporating Missing Data into Sample Size Calculations for CPMs

**Funding:** GPM, RW, GSC and RDR are supported by an MRC-NIHR Better Methods, Better Research grant [SS-Update: MR/Z503873/1]. GSC and RDR are NIHR Senior Investigators and RW, GSC and RDR are supported by the NIHR Birmingham Biomedical Research Centre at the University Hospitals Birmingham NHS Foundation Trust and the University of Birmingham. The views expressed are those of the author(s) and not necessarily those of the NHS, the NIHR or the Department of Health and Social Care.

**Competing Interests:** None

**Acknowledgements:** None

**Corresponding Author:**
Dr Glen Philip Martin
Senior Lecturer in Health Data Science
Vaughan House, University of Manchester, Manchester, M13 9GB, United Kingdom
Email: glen.martin@manchester.ac.uk

## Abstract

Clinical prediction models must be developed using sufficiently large datasets to minimise overfitting and ensure robust predictive performance. Existing sample size calculations assume complete predictor data for all included participants, yet missing values are common and may increase required sample sizes. This study aimed to quantify how missing predictor data and different imputation methods affect overfitting and model degradation, within datasets that adhere to current sample size criteria. We also aimed to explore how a general sample size framework based on anticipated posterior (sampling) distributions can be adapted to incorporate missing data assumptions and handling strategies. Using a simulation study, we found that in development data meeting current minimum sample size requirements, missing data reduced predictive performance, with expected calibration slopes frequently falling below the targeted value of 0.9. Increasing the required sample size to account for missing data reduced overfitting concerns, but the necessary inflation factor was context specific. In some scenarios, up to twice the minimum sample size was needed to achieve performance comparable to models developed using fully observed data. Expected value of perfect information calculations allowed quantification of the expected loss due to finite samples and missingness. Through two applied examples, we illustrate how embedding missing data assumptions and handling within the posterior sampling approach provides a principled way to determine required minimum sample sizes under missing data. Overall, missing predictor data increases minimum sample size requirements to develop stable and well-calibrated models. Our adaptations to recent posterior (sampling) sample size calculations offer a practical approach for incorporating missing data directly into sample size calculations.

# 1 Introduction

Clinical prediction models (CPMs) estimate an individual's probability of a future outcome using regression or machine learning models containing multiple predictors [1,2]. Such models must be developed using sufficiently large datasets to help ensure accurate and robust predictions [3]; insufficient sample size increases the risk of overfitting and degradation in model performance (e.g., miscalibration, and lower discrimination and clinical utility) [3–7]. To address this, several sample size calculations have been proposed for CPM development studies that aim to control overfitting, limit model degradation, and ensure precise estimation of key model parameters, such as the intercept [8–17]. Corresponding calculations also exist for CPM validation studies [18–23].

However, current sample size calculations implicitly assume that all individuals in the development dataset have completely observed values for all predictors being considered. In practice, missing predictor values are common throughout the CPM life-cycle and, if not handled appropriately, can introduce bias in both the derived model and estimates of predictive performance [24–31]. Although prior work has examined how missing data and imputation choices affect CPM development and performance [26–29,32–36], there has not been an investigation into how missing data assumptions and intended handling strategy should be incorporated into sample size calculations.

Since existing sample size calculations target the minimum sample sizes under fully observed data, we hypothesised that their direct application when missing data are anticipated in the development dataset may misalign the estimand of the calculation with the intended missing data handling strategy. In practice, this may translate into an insufficient effective sample size for model development, thereby increasing the risk of overfitting, model degradation, and loss of clinical utility. Therefore, this article aims to examine this issue and recommend adaptations to CPM sample size calculations that account for missing data concerns. Specifically, we extend the general sample size framework proposed by Riley et al. [11], which uses anticipated posterior (sampling) distributions of predictions under a given sample size and model development approach, to account for missing data assumptions and the planned handling strategy within the calculation. The Riley et al. [11] framework is flexible, applies to any modelling approach, and allows users to select targets for performance, assurance probabilities, stability, degradation and clinical utility. This flexibility means it can naturally accommodate missing data assumptions [37] and the intended handling strategy into the sample size calculation, as missing data can be introduced directly into the data generating mechanism used to derive the posterior distributions [11].

The objectives of this article are: (i) to quantify the impact of missing data assumptions and handling strategy on overfitting and model degradation when using datasets that meet current sample size criteria; and (ii) to demonstrate how the posterior sampling approach [11] can incorporate missing data assumptions and handling strategies into sample size calculations for CPM development. The remainder of the article summarises the posterior sampling procedure [11] and our adaptations (Section 2), evaluates the impact of missing data assumptions and handling strategy on sample size requirements through a simulation study (Section 3), and illustrates the proposed sample size approach through two practical examples (Section 4). For clarity, key terms used throughout this article are defined in **Box 1**.

**Box 1:** Definitions of key terms and concepts used throughout this article.

**Assurance probability**: The probability that a developed prediction model will achieve a specified level of performance within the target population. For example, a probability that the calibration slope will be between 0.9 and 1.1 [11].

**Calibration**: Agreement of the observed and predicted outcome values. For binary outcomes, this is the agreement between the observed and estimated outcome (event) probabilities. Assessed using calibration plots, and summary metrics such as observed:expected ratio, calibration intercept and calibration slope [2,38].

**Clinical utility**: The net benefit of using the model to aid clinical decision-making in individuals from a particular population. Specifically, where decision-making changes based on whether individuals have estimated probabilities above a particular probability threshold. Typically assessed using net benefit, which weights the benefits against potential harms, across a range of probability thresholds [39–42].

**Discrimination**: Ability of the prediction model to correctly differentiate those who experience the outcome and those who do not. That is, the probability that the model assigns a higher estimated probability to those who develop the outcome than those who do not develop the outcome. Assessed using the C-statistic (equivalent to the area under the receiver operating characteristics curve) for binary outcomes. [2,38].

**Expected Value of Perfect Information (EVPI)**: The expected loss in clinical utility (net benefit) attributable to uncertainty in the model's estimated probabilities, quantified as the difference between the net benefit achievable using the reference (true) model and that achieved using the developed model [43].

**Missing completely at random (MCAR)**: Where missingness is unrelated to observed and unobserved variables [37].

**Missing at Random (MAR)**: Where missingness is related to observed values of other variables but not directly to values of unobserved variables [37].

**Missing Not at Random (MNAR)**: Where missingness is related to unobserved values of other unobserved variables [37].

**Model degradation**: The difference in actual predictive performance of a reference (true) model that would be obtained under perfect information (i.e., infinite sample size) about predictor effects and the predictive performance of a developed model (based on a finite development dataset), for a target (deployment) population [11].

**Model Stability:** The degree to which a developed model (fitted on a fixed development dataset) is different (e.g., in terms of coefficients or selected predictor variables) to that obtained under a different random sample of the same size from the same target population [7,44–46].

**Overfitting**: Overfitting occurs when a model overly captures the idiosyncrasies of the development dataset, leading to overly extreme predictions in new individuals and thus worse predictive performance in the target population than observed in the development dataset. This arises when the model is too complex (in terms of number of parameters) relative to the available sample size [2,38] and when model development methods do not penalise (e.g., shrink) predictor effects for overfitting.

**Prediction Stability:** The extent that an individual's estimated probability differs across different models that have each been fit on different random samples of the same size from the same target population [44,45].

# 2 Adaptation of the general sample size framework based on anticipated posterior distributions to allow for missing data assumptions and handling strategy

**Figure 1** summarises the general sample size framework based on anticipated posterior (sampling) distributions that was proposed by Riley et al. [11]. The elements of **Figure 1** shown in bold represent our extensions to account for missingness and the chosen approach to handle missing data. Below, we expand on our adaptations; we refer readers to Riley et al. [11] for details of the full procedure.

## 2.1 Specify Missingness Assumptions: step 1.2 of Figure 1

The first adaptation requires researchers to specify assumptions about the missingness mechanism (MCAR, MAR, or MNAR [37]), expected missingness patterns (i.e., which predictor variables may be missing jointly), and the anticipated overall proportion of missing data. Such assumptions should be specified for both the development dataset and the large target population that are generated within the procedure (**Figure 1**). If missingness is expected to differ between development and deployment settings, then researchers can pre-specify different assumptions across each dataset. These assumptions should be informed by observed missingness patterns in real (or synthetic) data, if available, [11] ideally with stakeholders who understand the data observation processes. Graphical representations of missing data may help with informing and articulating these assumptions [47].

Researchers should decide whether they are targeting a CPM that will be ultimately deployed where all individuals have fully observed predictor variables (sometimes called *ideal performance* [31]), or a CPM that will be deployed where some individuals have missing predictor variables (sometimes called *pragmatic performance* [31]). For instance, we might be in a situation where missing data will occur within the development dataset, but where all data will be required to be observed when making predictions during deployment of the model.

## 2.2 Inducing Missingness in the Simulated Datasets: steps 2.1b and 2.2b of Figure 1

After specifying the missingness assumptions and simulating both a large target population and development dataset (see steps 2.1 and 2.2a), missing data are induced into the simulated datasets. If the CPM is anticipated to be deployed not allowing for missing data (i.e., *ideal performance* [31]), then missingness is induced only in the development dataset. In contrast, if the CPM is anticipated to be deployed in the presence of missing data (i.e., *pragmatic performance* [31]), then missingness is induced in both datasets to reflect real deployment conditions. Embedding missingness directly in the data-generating mechanism ensures that performance metrics estimated include the combined effects of sampling variation and missing data. Inducing missing data into the datasets can be achieved using fully conditional specification in multivariate imputation [48]; the *ampute* function from the *mice* R package implements this [49].

### 2.3 Handle Missing Data During CPM Development and Validation: stages 3 and 4 of Figure 1

The planned method of handling missing data during CPM development (e.g., multiple imputation) is then applied to the simulated development dataset. A CPM would then be developed using all the planned analytical steps (such as variable selection) on the imputed development data.

If targeting *pragmatic performance* [31], the missing data within the large target population would then be imputed using the same imputation models learned during development. Re-estimating imputation models in the target data should be avoided because doing so produces biased estimates of predictive performance [27,28].

The developed CPM is then evaluated on the large target population, either the fully observed version for *ideal performance* [31] or the above imputed version for *pragmatic performance* [31]. Here, any performance metrics can be used to quantify predictive performance [50]. Researchers should also quantify the predictive performance of the reference (true) model in the respective large target population dataset, to allow calculation of degradation in performance (see **Box 1**).

### 2.4 Repeat the Procedure, and increase $N_{dev}$ as required: stage 5

As in Riley et al. [11], this procedure is then repeated many times (e.g., to produce 1000 samples), to obtain anticipated posterior distributions of the performance metrics (e.g., calibration, discrimination, and net benefit). Researchers then evaluate whether the development sample size ($N_{dev}$) (e.g., the sample size suggested by quick closed-form criteria may be used as a starting point [10]) meets their context-specific performance targets (e.g., expected calibration, assurance probability, degradation, or clinical utility). Increasing the development sample size ($N_{dev}$) and repeating the entire procedure, as required, allows identification of the minimum sample size needed.

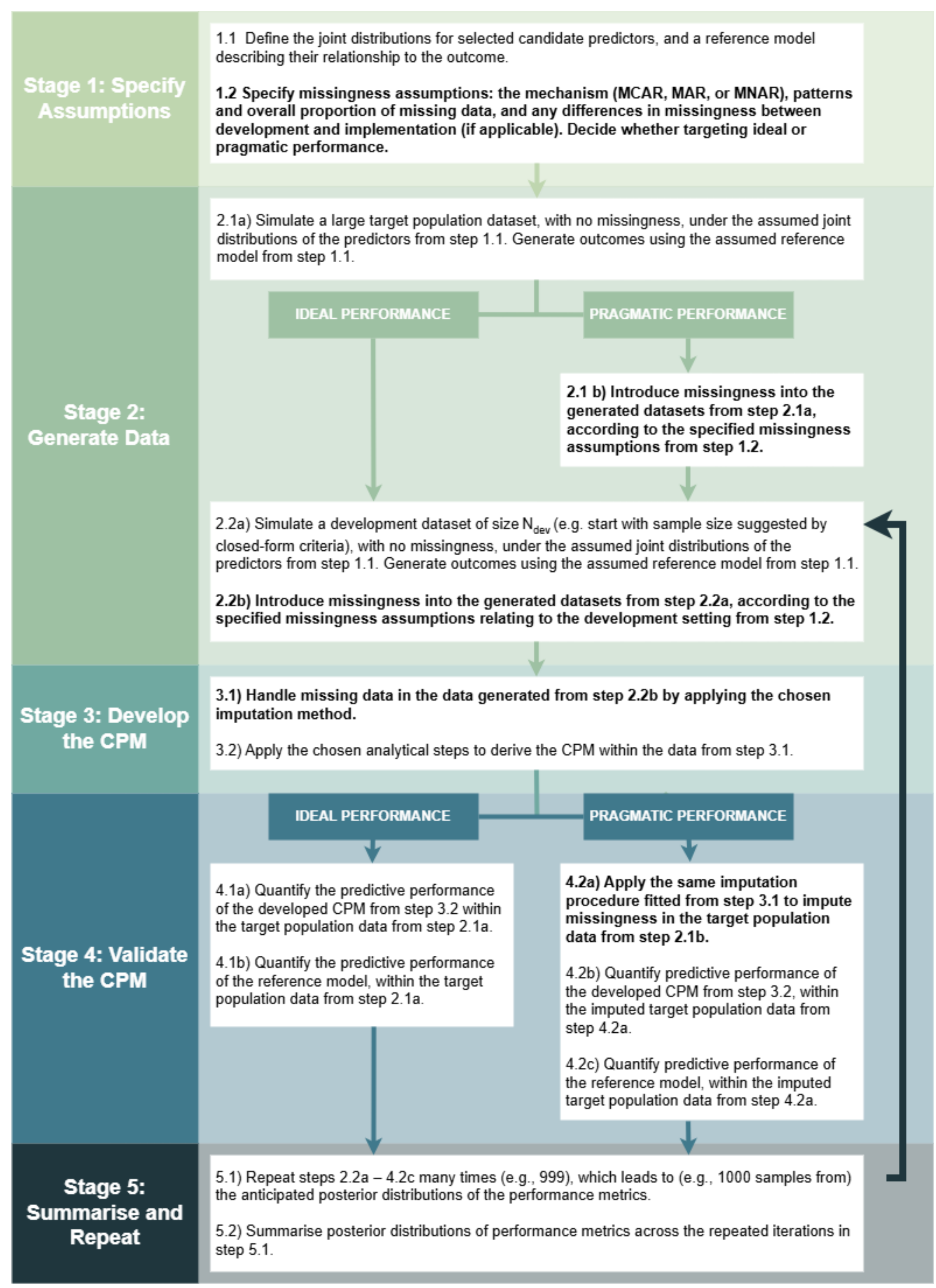

**Figure 1:** Overview of the general sample size procedure based on anticipating posterior distributions proposed by Riley et al., [11] adapted to include our extension (items in bold) to incorporate missing data handling. Repeat the entire procedure, as required, across increased assumed sample sizes ($N_{dev}$ in step 2.2a) until the desired level of performance from step 5.3 (which will be context-dependent and chosen by the user) is achieved.

# 3 Simulation Study

Although Section 2 provides a framework for embedding missing data assumptions and handling strategy within sample size calculations for CPM development, the broader question is whether this is necessary compared to researchers simply assuming complete data for the calculation, and indeed whether simpler closed-form solutions would suffice [10]. In this section we describe a simulation study, following the ADEMP framework [51], to examine this question.

## 3.1 Simulation Aim

The aim of this simulation study was to evaluate how missing data and its handling, impacts the overfitting, model degradation and expected value of perfect information (EVPI) when developing a CPM across a range of samples sizes. We had particular interest in examining the impact when using a minimum sample size calculated by closed-form sample size criteria [8–10].

## 3.2 Data Generating Model

The design of the simulation data generating models mimic stages 1 and 2 of **Figure 1**. Specifically, we begin by generating a large target population of size $N = 500{,}000$. We assume (without loss of generality) we have a binary outcome event, $\boldsymbol{Y}$, that we wish to predict conditional on ten predictor variables, denoted $\boldsymbol{X}_p$ for $p = 1, \dots, 10$. The predictor variables were drawn from a multivariate normal distribution such that

$$\boldsymbol{X} \sim MVN(\mathbf{0}, \boldsymbol{\Sigma})$$

where $\boldsymbol{\Sigma}$ is a block-diagonal covariance matrix. Predictor variables were grouped into pairs, with each pair potentially correlated at values $\rho \in \{0, 0.5, 0.75\}$, depending on simulation scenario. Each pair was uncorrelated with other variables. For example, $\boldsymbol{X_1}$ and $\boldsymbol{X}_2$ could be correlated but uncorrelated with other variables; similarly, $\boldsymbol{X}_3$ and $\boldsymbol{X}_4$ formed another potentially correlated pair, and so forth. We generate the outcome according to a logistic regression reference model such that

$$P(y_i = 1) = expit(\beta_0 + \beta_1 x_{i,1} + \cdots + \beta_{10} x_{i,10})$$

Here, the coefficients $\beta_1, \dots, \beta_{10}$ represent 'true' log-odds ratios, which were varied across two simulation scenarios. First, a scenario where all predictor variables were associated with the outcome, with $\beta_1 = 0.5, \beta_2 = 0.2, \beta_3 = 0.3, \beta_4 = 0.1, \beta_5 = 0.5, \beta_6 = 0.2, \beta_7 = 0.3, \beta_8 = 0.05, \beta_9 = 0.1, \beta_{10} = 0.15$. Second, a scenario where five of the predictor variables were 'noise' terms; that is, $\beta_1 = 0.5, \beta_2 = 0, \beta_3 = 0.3, \beta_4 = 0, \beta_5 = 0.5, \beta_6 = 0, \beta_7 = 0.3, \beta_8 = 0, \beta_9 = 0.1, \beta_{10} = 0$. These values were selected to be largely representative of predictor effects in published CPMs and allowed us to explore scenarios where two predictor variables might be correlated, but only one – or both – might be associated with the outcome (depending on the scenario of $\beta_1, \dots, \beta_{10}$ and $\rho$). This reflects realistic complexities that may occur in health data. In all cases, $\beta_0$ was selected in each simulation to give an overall outcome prevalence of 20% (without loss of generality since we report all results relative to minimum closed-form sample size criteria [8–10] that account for prevalence of the outcome).

We then used the closed-form sample size criteria [9] to estimate an initial minimum sample size ($N_{min}$) , using the *pmsampsize* package in R [52,53]. For these calculations, the number of predictor parameters was set to 10, the outcome prevalence set to 20%, the desired level of shrinkage set at 0.9, and we assumed a conservative Nagelkerke's $R^2$ of 0.15 [9]. Based on $N_{min}$, we generated development data of size $N_{dev} = \delta \times N_{min}$, using the same data generating model as that described above for the large target population. The value of $\delta$ was varied across simulation scenarios with values $\{1, 1.25, 1.5, 1.75, 2\}$, allowing us to explore the impact of increasing development sample size beyond that obtained by the closed-form criteria. For every simulation scenario, we generated 100 different development datasets, each of size $N_{dev}$. The choice of 100 iterations was chosen for computational reasons across the large number of scenarios we considered.

Missing data were then introduced into the large target population, and into each of the 100 development datasets per simulation scenario. Missingness was assumed to be either MAR or MNAR mechanisms (depending on simulation scenario), with the overall proportion of missingness, $\pi_{\text{miss}}$, varying by scenario ($\pi_{miss} \in \{0.1, 0.2, 0.4, 0.6\}$). For this simulation, we assumed the missingness mechanism was the same between development and deployment (target population dataset); as outlined in Section 2, this could be relaxed in practice. Missingness was applied across all 10 predictor variables using fully conditional specification in multivariate imputation [48]. Specifically, for MAR, missingness in each variable $X_p$ for $p \in [1,10]$ depended on all other predictor variables other than $X_p$, with each variable having equal weight to the missingness probability of $X_p$. This was similar for MNAR, except that missingness in $X_p$ also depended on that variable itself in addition to the other predictor variables.

This data generating process was repeated across all combinations of the simulation parameter values (**Table 1**), resulting in 240 simulation scenarios. The simulation scenarios were selected based on our prior experience of analysing real health data.

**Table 1:** Summary of the simulation parameters and the values considered in the simulation, giving 240 simulation scenarios across all combinations of them.

| **Simulation Parameter** | **Values considered** |
|---|---|
| Predictor weights ($\beta_1, \ldots, \beta_{10}$) | $\beta_1 = 0.5, \beta_2 = 0.2, \beta_3 = 0.3, \beta_4 = 0.1, \beta_5 = 0.5, \beta_6 = 0.2, \beta_7 = 0.3, \beta_8 = 0.05, \beta_9 = 0.1, \beta_{10} = 0.15$<br>or<br>$\beta_1 = 0.5, \beta_2 = 0, \beta_3 = 0.3, \beta_4 = 0, \beta_5 = 0.5, \beta_6 = 0, \beta_7 = 0.3, \beta_8 = 0, \beta_9 = 0.1, \beta_{10} = 0$ |
| Prevalence of the binary outcome (controlled by $\beta_0$) | 20% |
| Correlation between predictors ($\rho$) | 0, 0.5 or 0.75 |
| 'Multiplier' to the minimum sample size to derive the size of development dataset ($\delta$) | $1, 1.25, 1.5, 1.75$ or $2$ |
| Missingness mechanism | MAR or MNAR |

| Overall proportion of individuals in the development sample who have missing values in at least one predictor variable ($\pi_{miss}$) | $0.1, 0.2, 0.4$ or $0.6$ |
|---|---|

### 3.3 Estimand

The estimand is the estimated probability of the outcome, conditional on the covariates, within the large overarching target population. The metrics used to evaluate a given CPMs performance at estimating this estimand are described in Section 3.5.

### 3.4 Methods being Compared

Within each of the 240 simulation scenarios, we applied the following imputation methods to each of the 100 generated development datasets: complete case analysis, mean imputation, single regression imputation, random forest imputation [34] and multiple imputation by chained equations. For single regression imputation and multiple imputation, linear regression was used as the underlying imputation models, and for multiple imputation we generated 20 imputed datasets by chained equations. For single regression imputation and random forest imputation, the imputation models for each missing predictor variable included all other predictor variables as covariates but not the outcome [54]. For multiple imputation, the imputation models included all predictor variables and the outcome as covariates [54].

To each imputed development dataset, we developed three CPMs using: (i) unpenalised (maximum likelihood) logistic regression, (ii) unpenalised logistic regression with predictor selection using Akaike Information Criterion (AIC) backwards selection, and (iii) logistic regression with a LASSO penalty. All models considered the ten candidate predictor variables $X_1, \dots, X_{10}$. For comparison, these three models were also fit to the fully observed version of the development dataset, prior to introducing missing values to quantify the impact of missingness.

### 3.5 Performance Metrics

We quantified and compared the predictive performance of each fitted CPM within the fully observed target population and the target population with missing data, reflecting performance when CPM is anticipated to be deployed not allowing for, and allowing for, missing data, respectively (**Figure 1**). In addition, we quantified the performance of the reference model (i.e., if we knew the true predictions for everyone).

Our primary performance metric was the calibration slope, serving as an estimate of the shrinkage factor and indicating the degree of overfitting. This was calculated by fitting a logistic regression model to the observed outcomes in the target population, using the logit of the estimated probabilities from the CPM under evaluation as the sole covariate. We calculated the median calibration slope across the 100 repeats of each of the 240 simulation scenarios and calculated the assurance probability of the calibration slope being between 0.9 and 1.1 (i.e., proportion of the 100 repeats for each scenario where the calibration slope lied between these values).

As secondary performance metrics, we calculated the degradation in the developed CPM's predictive performance (observed-to-expected ratio, calibration intercept, and C-statistic), as defined in **Box 1**. Additionally, we estimated the EVPI for each developed CPM when applied within the target population (**Box 1**), across a range of threshold values from 0 to 1, and presented the results graphically [43,55].

### 3.6 Software

The simulation was implemented using R version 4.5.1, along with the tidyverse [56], glmnet [57], pmsampsize [58], mice [49], missForestPredict [34], and predRupdate [59] packages. The simulation code and results can be found on GitHub (https://github.com/GlenMartin31/MissingDataSampleSize).

### 3.7 Simulation Study Results

Across all simulation scenarios, the closed-form minimum sample size was 897, driven by a target calibration slope of 0.9. Accordingly, $N_{dev}$ took values of 897, 1122, 1346, 1570 and 1794 for $\delta = \{1, 1.25, 1.5, 1.75, 2\}$, respectively. Below, we summarise the main findings for scenarios with moderate correlation between the predictor variables ($\rho = 0.5$), with data MAR, and where all predictor variables are associated with the outcome; other scenarios showed similar findings (see **Supplementary Figures 1 – 5**).

#### *3.7.1 Missing data increases risk of overfitting at the closed-form sample size*

The median calibration slope of the unpenalised CPM fitted to fully observed development data was close to 0.9 when $N_{dev} = N_{min}$ (**Figure 2**). In contrast, as the level of missingness increased, the median calibration slope dropped below 0.9 across imputation methods, with complete case analysis (i.e., simply excluding any participants with missing data) particularly prone to overfitting, as one might expect. Model performance declined further when assessing *pragmatic performance* [31], where missingness was also present in the target population (**Figure 2: Panel B**), reflecting the additional uncertainty introduced when predictions must be made for individuals with incomplete data [27,31]. Missing data also increased instability at the closed-form sample size $N_{min}$: the assurance probability of the calibration slope being between 0.9 and 1.1 fell from 0.53 under fully observed data to below 0.4 when 40–60% of individuals had missing predictor values (**Figure 3**).

#### *3.7.2 Greater minimum sample sizes may be required if the development data are expected to contain missing values*

The median calibration slope of the unpenalised CPMs for each imputation method tended towards 0.9 as the sample size increased (**Figure 2**). A similar pattern was observed in the degradation of other performance metrics (**Supplementary Figure 2**). For example, with 40% missingness, increasing the sample size by 25% (i.e., $\delta = 1.25$) resulted in the degradation of the C-statistic for models developed under regression imputation, multiple imputation or random forest imputation being similar to that observed under fully observed development data.

Sample sizes up to twice the closed-form criteria were sometimes necessary to achieve model stability and assurance probabilities comparable to those observed under fully observed data (**Figure 3**). For instance, with 60% missingness, the CPM developed under multiple imputation achieved assurance probabilities similar to the fully observed data CPM only when $\delta = 2$.

### 3.7.3 *Expected Value of Perfect Information (EVPI) can quantify the expected loss (in net benefit) due to the finite development sample and due to missing data*

The simulation-based sample size procedure allowed calculation of EVPI, which quantifies the loss in decision-making utility attributable to uncertainty from both finite sample sizes and missing data (**Figure 4**). As the proportion of missing data increased, the EVPI curves for the CPMs developed under imputed datasets diverged further from those for CPMs developed using fully observed data. Increasing the sample size reduced this gap, with EVPI curves for CPMs developed on imputed data converging toward those of the CPM developed on fully observed data as $\delta$ increased (i.e., comparing down the columns of plots in **Figure 4**).

### 3.7.4 *Penalisation methods improve overfitting on average, but instability remains a concern*

CPMs fit under LASSO penalised likelihood had a median calibration slope above 0.9 across all levels of missing data we considered, even at the closed-form sample sizes (**Figure 2**). However, the assurance probabilities indicated instability similar to unpenalised models (**Figure 3**). This highlights that while penalisation can reduce average overfitting, it does not fully mitigate the uncertainty introduced by missing data, reinforcing the value of evaluating assurance-based metrics within the general sample size framework [11].

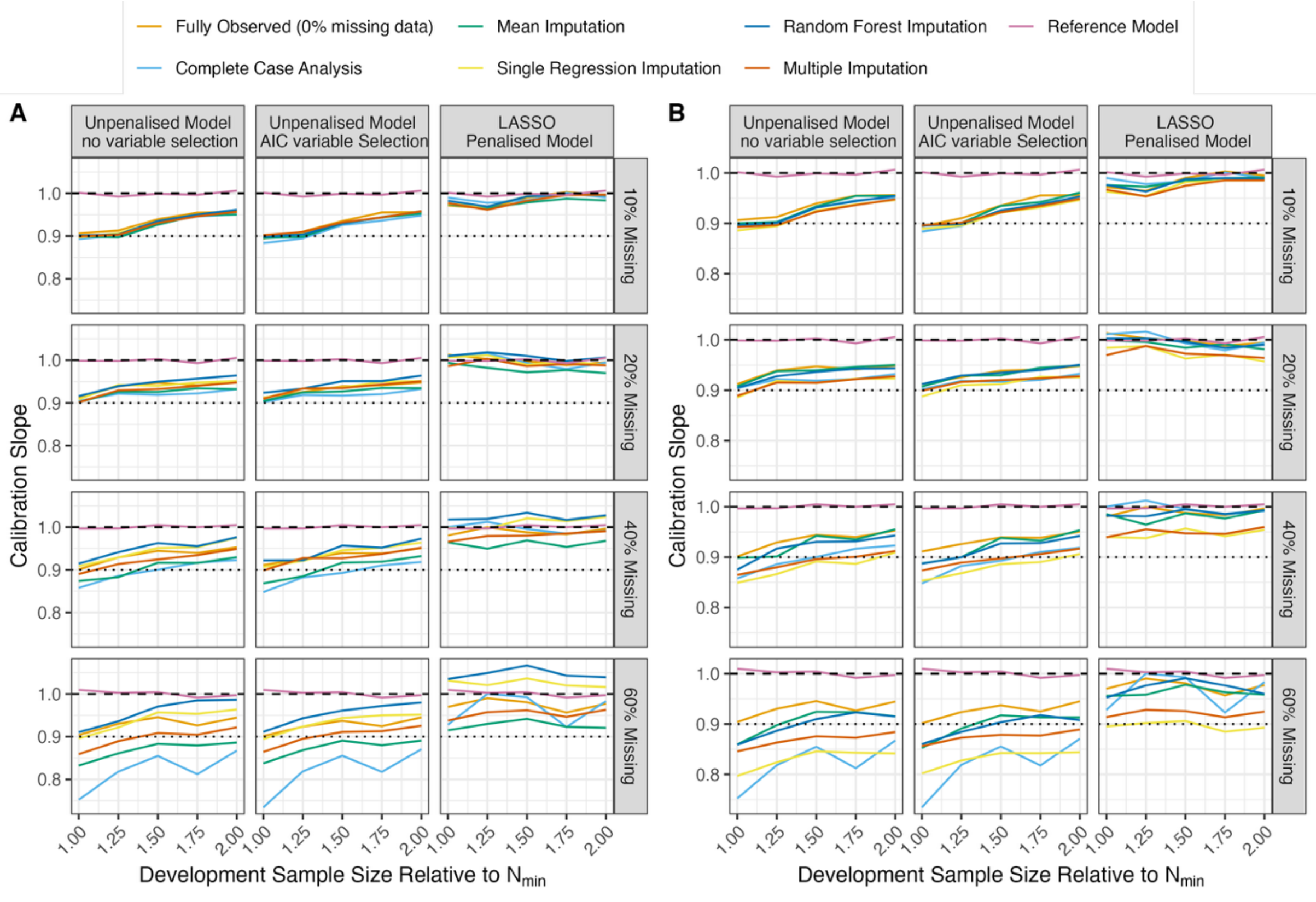

**Figure 2:** Median calibration slope for each model across different sample sizes relative to the closed-form minimum sample size calculation, and different levels of missing data. Panel (A) shows the results estimated within the fully observed target population and Panel (B) shows the results estimated within the imputed version of the target population (using the imputation procedure corresponding to each CPM). Results are for cases where data were missing at random (MAR), for $\rho = 0.5$ and where all predictor variables are associated with the outcome. The dotted horizonal line shows that targeted value of 0.9, and the dashed horizontal line shows a perfect calibration slope of 1.

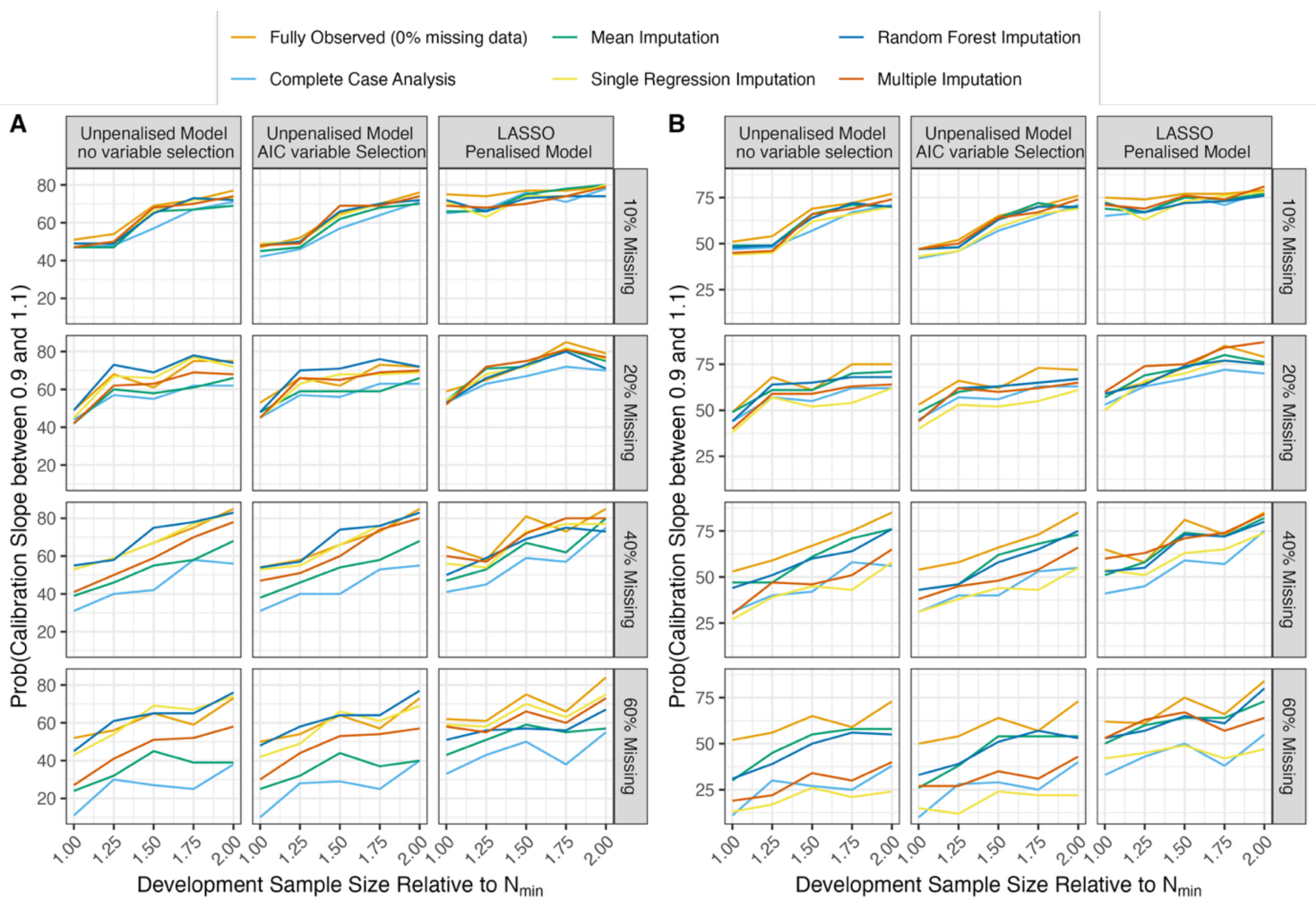

**Figure 3:** Assurance probability of the calibration slope being between 0.9 and 1.1 for each model across different sample sizes relative to the closed-form minimum sample size calculation, and different levels of missing data. Panel (A) shows the results estimated within the fully observed target population and Panel (B) shows the results estimated within the imputed version of the target population (using the imputation procedure corresponding to each CPM). Results are for cases where data were missing at random (MAR), for $\rho = 0.5$ and where all predictor variables are associated with the outcome.

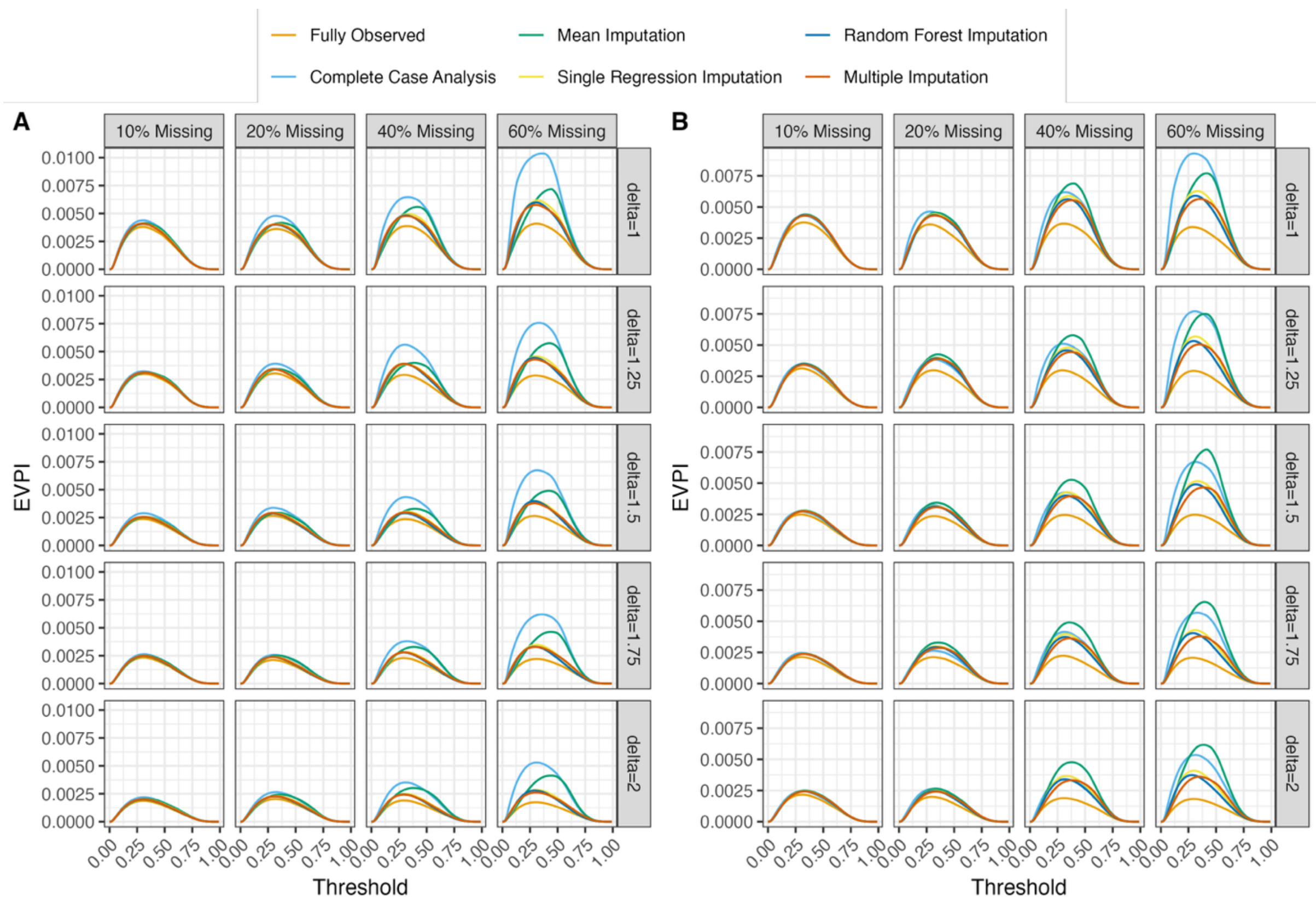

**Figure 4:** Expected Value of Perfect Information (EVPI) for each model's net benefit (with maximum net benefit defined by the reference model) across different sample sizes relative to the closed-form minimum sample size calculation (rows of each panel; delta), and different levels of missing data. Panel (A) shows the results when data were MAR and Panel (B) shows the results when data were MNAR. Results are for cases where $\rho = 0.5$, where the model was fitted using unpenalised likelihood, and where all predictor variables are associated with the outcome.

## 4 Practical Examples

In this section, we illustrate the practical application of the simulation-based sample size procedure (**Figure 1**). We imagine we are designing a new study to develop a logistic regression CPM for in-hospital mortality among patients admitted to an intensive care unit (ICU) using predictor variables available at the end of the first 24 hours of admission. We present two examples, differing in the availability of data to help us make the required assumptions of the sample size procedure. For both, the CPM will be deployed in settings where all predictor values must be observed, but missing data are anticipated in the development dataset. The same principles would apply if the CPM was to be deployed allowing for missing data (**Figure 1**).

Across both examples, we plan to consider 21 candidate predictor parameters, including age (categorical, 7 levels), gender (categorical, 2 levels), ethnicity (categorical, 3 levels), admission type (categorical, 2 levels), and several laboratory markers from the first 24 hours (all continuous): bicarbonate, creatinine, chloride, haemoglobin, platelet count, potassium, partial thromboplastin time, international normalised ratio, prothrombin time, blood urea nitrogen, and white blood cell count.

### 4.1 Example 1: Individual Participant Data (IPD) Available

In this first example, we assume we have access to the Medical Information Mart for Intensive Care III (MIMIC-III) database [60] that we will use to help inform the assumptions of calculating the minimum sample size of our new study. MIMIC-III contains detailed clinical data from ICU admissions at the Beth Israel Deaconess Medical Center between 2001 and 2012. These data allow us to approximate: (i) the joint distributions of the candidate predictor variables, (ii) the reference (true) model, and (iii) the magnitude and patterns of missingness that might be observed in the new data. The steps of **Figure 1** then proceed as follows.

**Step 1.1:** We estimated the joint distributions of predictor variables using the *synthpop* R package [61] applied to the MIMIC-III dataset. The reference (true) model was taken from a previously published logistic regression model for in-hospital mortality developed using MIMIC-III [7] (**Supplementary Table 1**).

**Step 1.2:** The overall level of missingness across the MIMIC-III dataset was 22%, so we assume the same will be observed in our new study. The missingness mechanism and distributions were estimated using the MIMIC-III dataset, using the procedure outlined in step 2.2b below. This gives the anticipated missingness magnitude and mechanisms for new data collection for the new study.

**Step 2.1a:** We generate a large (N=500,000) target population dataset using the *synthpop* R package [61], based on the observed joint distributions in the MIMIC-III dataset. Here, only observed data were included, meaning the generated synthetic data was 'fully observed'. Outcomes were generated from a binomial distribution using the 'true' probabilities from the model in step 1.1.

**Step 2.1b:** Since we are targeting a CPM where all data will be required upon deployment, we skip this step and retain the 'fully observed' version of the large target population generated from step 2.1a.

**Step 2.2a:** A separate development dataset was generated under the same process as described in step 2.1a above. The closed-form procedure of Riley et al. [9] was taken as the initial value of $N_{dev}$, using a previously reported C-statistic of 0.73 [7], a target shrinkage factor of 0.9 and 21 candidate predictor parameters.

**Step 2.2b:** To impose realistic missingness patterns on the fully observed development dataset from Step 2.2a, we generated a second synthetic dataset of identical size ($N_{dev}$) using *synthpop* on the MIMIC-III dataset, this time allowing missing values to occur. We overlaid the resulting missingness pattern from this generated dataset onto the dataset from Step 2.2a. Specifically, for each individual and variable, the data value from the Step 2.2a dataset was retained if the corresponding data in that from Step 2.2b was observed or was set to missing if it was missing in the Step 2.2b dataset.

**Steps 3.1 & 3.2:** We used the data from step 2.2b to fit an unpenalised logistic regression model with all 21 candidate predictor parameters, and no variable selection. Missing data were handled using four alternative strategies: complete case analysis, random forest imputation, single regression imputation, and multiple imputation by chained equations. Although in practice a single imputation strategy would be pre-specified, we considered multiple approaches here simply to illustrate the procedure for different imputation methods. For comparison, we also fitted a CPM on the fully observed version of the development dataset (i.e., the data from step 2.2a).

**Steps 4.1a & 4.1b:** We evaluate each developed CPM in the large target population dataset obtained from step 2.1a. Predictive performance was summarised using the calibration slope and C-statistic. These metrics were chosen simply for illustration; as stated in Riley et al., [11] users can choose any performance metric they wish for this step.

**Step 5.1**: The above steps were repeated 100 times to characterise the distribution of performance metrics under each imputation strategy. We calculated assurance probabilities of the calibration slope, defined as the proportion of the 100 iterations that the calibration slope for a given model was between 0.9 and 1.1.

**Steps 5.2 & 5.3**: The initial value of $N_{dev}$ was calculated to be 2836. At this sample size, we found that the median calibration slope was 0.85, 0.89, 0.89 and 0.88 for the model developed under complete case analysis, random forest imputation, single regression imputation and multiple imputation, respectively (**Figure 5**). For the CPM fit on the fully observed version of the development dataset, the median calibration slope was 0.89. Whilst these average values quite closely agree with the targeted level of overfitting (0.9), the assurance probability was below 45% for all models. The median C-statistic of all models apart from complete case analysis was similar at 0.721 (**Supplementary Figure 6**). Note, the level of missing data was 22% in the dataset, so the similarity in results between the methods aligns with the findings from our simulation study.

**Step 5.4**: Given the relatively low assurance probability, we re-ran the procedure increasing the value of $N_{dev}$ to be 1.5 times higher ($N_{dev} = 4254$) and 2 times higher ($N_{dev} = 5672$). At samples sizes 1.5 times higher than closed-form solutions, the median calibration slope for all models was above 0.9, with assurance probabilities ranging from 56% for complete case analysis to 73% for random forest imputation (**Figure 5**). At sample sizes 2 times higher than closed-form solutions, the median calibration slope was above 0.94 for all CPMs, with assurance probabilities above 70% for all models except complete case analysis. It would be context-dependent whether we would then choose $N_{dev} = 4254$ or $N_{dev} = 5672$ for the design of our new study.

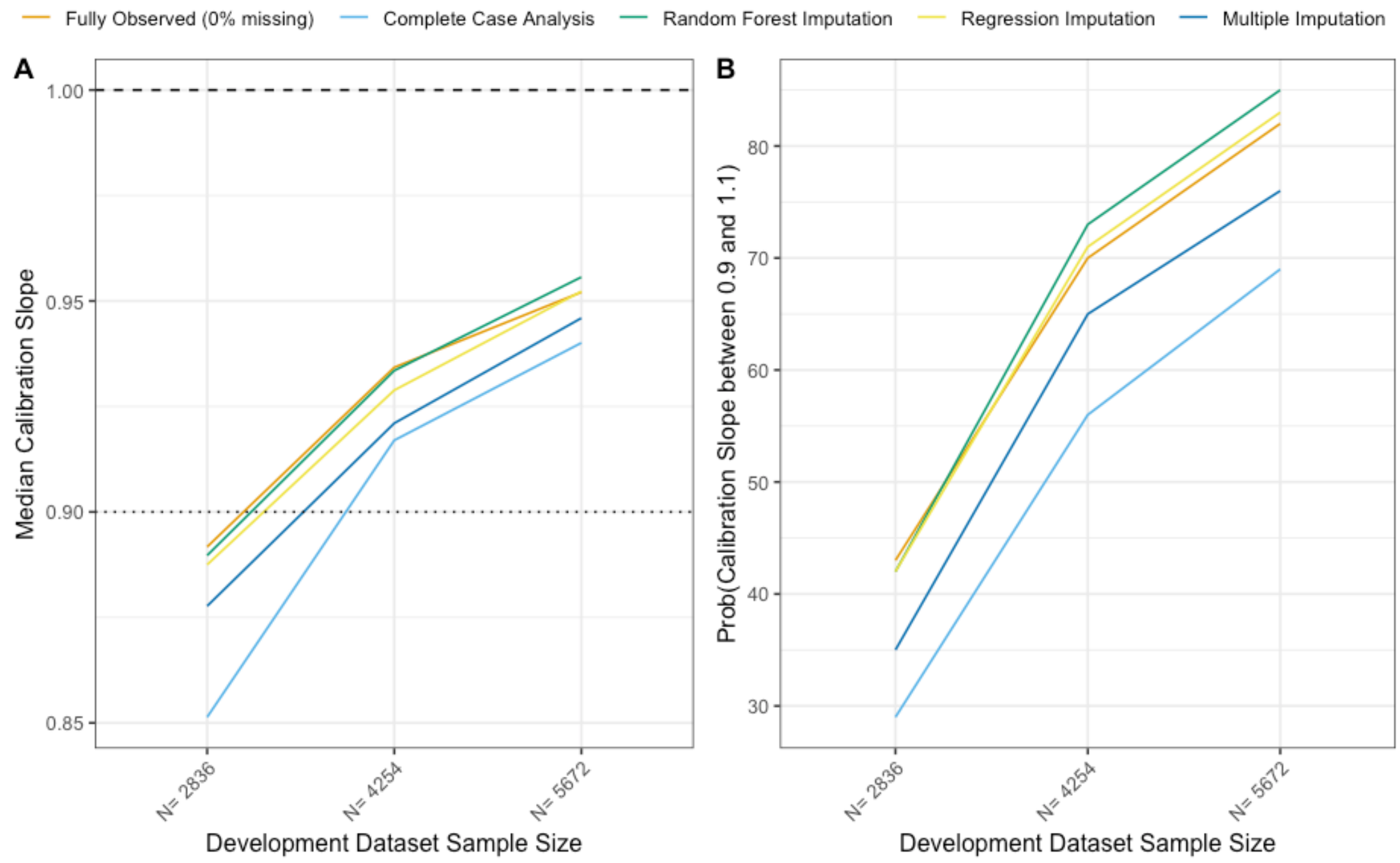

**Figure 5:** Median calibration slope (Panel A) and assurance probability (Panel B) for each model across different sample sizes for practical example 1.

### 4.2 Example 2: No Individual Participant Data (IPD) Available

Our second example again considers a situation where we are designing a new study to develop a prediction model for mortality prediction on intensive care. However, this time we assume we have no access to existing data to help inform the assumptions within the proposed sample size procedure (**Figure 1**). The full details of this example are outlined in the supplementary materials. In short, we assumed MNAR across all combinations of missingness pattern. To induce missingness into the datasets, we used fully conditional specification in multivariate imputation [48], implemented using the *ampute* function within the *mice* R package [49]. Overall, we found similar results to those described above for example 1. For example, at the initial sample size of 2836, the median calibration slopes were 0.83, 0.87, 0.87 and 0.86 for complete case analysis, random forest imputation, single regression imputation and multiple imputation, respectively (**Supplementary Figure 7**). Again, the assurance probabilities were all <40% at the initial sample size. Increasing the sample size to 4254 resulted in median calibration slopes above 0.9, with assurance probabilities >50%, for all models expect complete case analysis (**Supplementary Figure 7**).

## 5 Discussion

This study has shown that when development datasets contain missing predictor values, closed-form sample size calculations [8–10] (which assume complete data) underestimate the sample sizes needed to achieve stable and reliable predictive performance. We have extended the general simulation-based framework of Riley et al.[11] to incorporate assumptions about missing data mechanisms and the analyst's planned approach to handle missingness (**Figure 1**). Importantly, the procedure can be used regardless of the planned modelling strategy or the planned method of handling the missing data.

Our findings build upon the growing research on sample size requirements for CPM development [5,8–17,62] and represent the first explicit assessment of how missing data influences these requirements. Since uncertainty introduced by missingness (and its handling) propagates through the entire modelling pipeline, it is unsurprising that missing data substantially reduced predictive performance compared to fully observed data. When datasets contained moderate or high levels of missingness, calibration slopes were often below the targeted expected value of 0.9. Increasing the sample size reduced overfitting; in some scenarios, sample sizes up to twice the closed-form minimum were required. This underscores the need to consider missingness during the design phase of CPM development, and our proposed adaptation to the general framework by Riley et al. [11] provides a way for researchers to do this.

Although the procedure requires assumptions about the predictor variable joint distributions, missingness mechanisms and missingness patterns, all sample size calculations (whether for a prediction modelling study or otherwise) rely on assumptions. The key is to articulate those assumptions clearly and ground them in available data, literature, and expert knowledge. In situations where these assumptions cannot reasonably be specified (e.g., if there is a lack of any prior information), one pragmatic alternative is to combine closed-form calculations [8,9] with penalised regression (e.g., LASSO). Our simulations suggest that this controls the level of overfitting on average, although assurance probabilities for a single development dataset may still be quite low [11,63]. Indeed, penalisation or shrinkage methods can underestimate the required level of shrinkage in some situations [63], which can lead to unstable prediction models [6,7,44]. Recent adaptations to the closed-form solutions by Pavlou et al. [17] may help.

An alternative strategy in the absence of information to inform the simulation-based sample size procedure, is to use a conservative rule-of-thumb to inflate the closed-form calculations [8,9]. Our simulations indicate that inflating the closed-form sample size calculations [8,9] by a factor of $1/(1-\pi_{miss})$, where $\pi_{miss}$ is the overall proportion of missing data in the dataset, provides an approximate adjustment for the average impact of missing data. However, the necessary inflation factor is context specific (including the choice of model development approach), so the proposed simulation-based procedure (**Figure 1**) remains the preferred method whenever feasible.

While our results indicated that different imputation methods required different sample sizes, we emphasise that the choice of method to handle missing data should reflect how missingness will be handled when the CPM is deployed, not which approach seems to meet a given sample size requirement. Ensuring consistency in missing-data handling across development, validation, and deployment is crucial [24,26–28,35]. If required sample sizes for the selected missing data handling approach exceeds the data available, then this calls for a pause in modelling until more data can be obtained or to adjust modelling complexity (e.g., reduce number of candidate predictor variables). We particularly encourage the use of EVPI as a way of quantifying the expected loss (in model decision-making utility) due to both finite sample and missing data, as illustrated in **Figure 4**.

Several limitations warrant consideration. First, although we examined 240 simulation scenarios, they do not cover all possible data structures; our choices were intended to give indication of the impact of missing data across a range of data structures. Second, specifying the missingness mechanisms can be challenging, and misspecification may impact minimum sample size estimates. Future work should explore how best to elicit or define these assumptions, and to assess the impact of assumption misspecification. Finally, we focussed on missing predictor variables, with non-hierarchical data. Many applications involve missing outcomes or hierarchical structures (e.g., temporally repeated measures per person), which both introduce further complexities requiring future research.

In conclusion, missing predictor data increases the minimum sample size required to have assurance that a developed CPM will be stable, well-calibrated and clinically useful. Since missingness is common across the CPM lifecycle, explicitly accounting for this during the study design phase is essential. Our extended simulation-based framework provides a practical and principled way to do this based on sensible assumptions about the type and magnitude of missing predictor data and the planned handling strategy.

# Supplementary Material: Incorporating Missing Data Considerations into Sample Size Calculations for Developing Clinical Prediction Models

Table of Contents

# 7 Supplementary Simulation Study Results

This section presents supplementary figures and results from our simulation study described within the main paper.

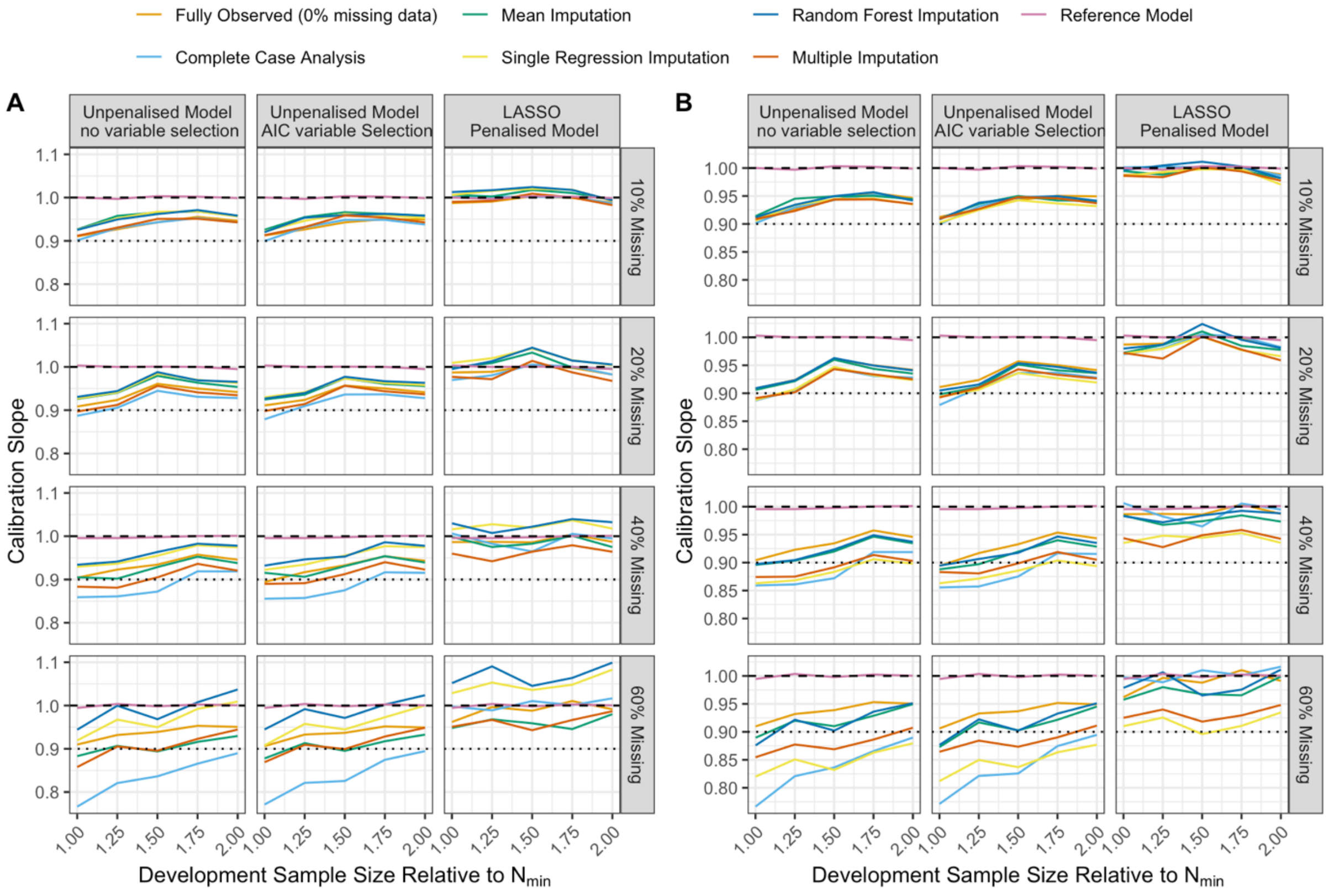

**Supplementary Figure 1:** Median calibration slope for each model across different sample sizes relative to the closed-form minimum sample size calculation, and different levels of missing data. Panel (A) shows the results estimated within the fully observed target population and Panel (B) shows the results estimated within the imputed version of the target population (using the imputation procedure corresponding to each CPM). Results are for cases where data were missing not at random (MNAR), for $\rho = 0.5$ and where all predictor variables are associated with the outcome. The dotted horizonal line shows that targeted value of 0.9, and the dashed horizontal line shows a perfect calibration slope of 1.

**MAR**

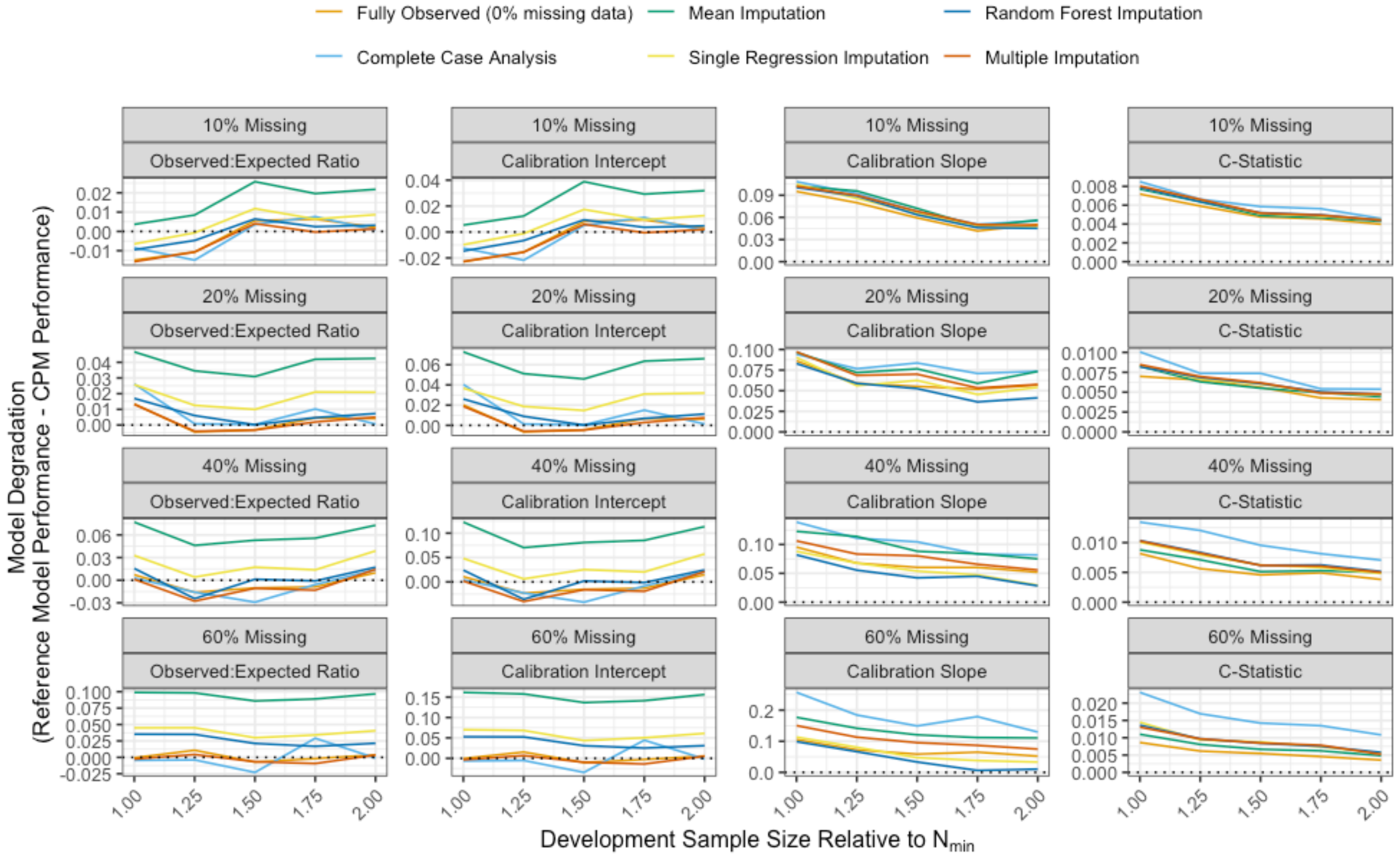

**MNAR**

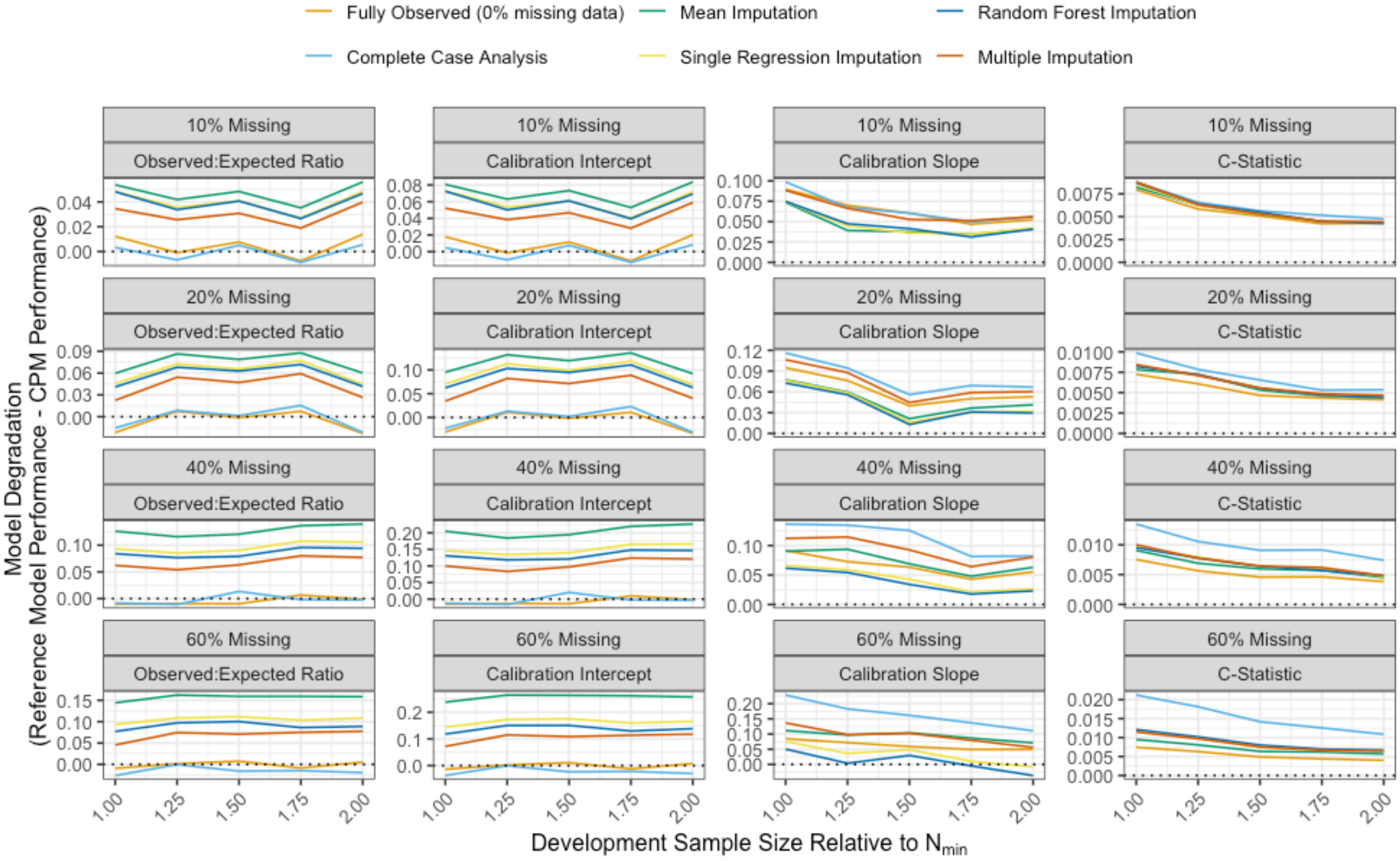

**Supplementary Figure 2:** Model degradation for each model across different sample sizes relative to the closed-form minimum sample size calculation, and different levels of missing data. **Top panel:** shows the results when data are MAR. **Bottom panel:**

shows the results when data are MNAR. Results are where $\rho = 0.5$, for the unpenalised CPM, and where all predictor variables are associated with the outcome.

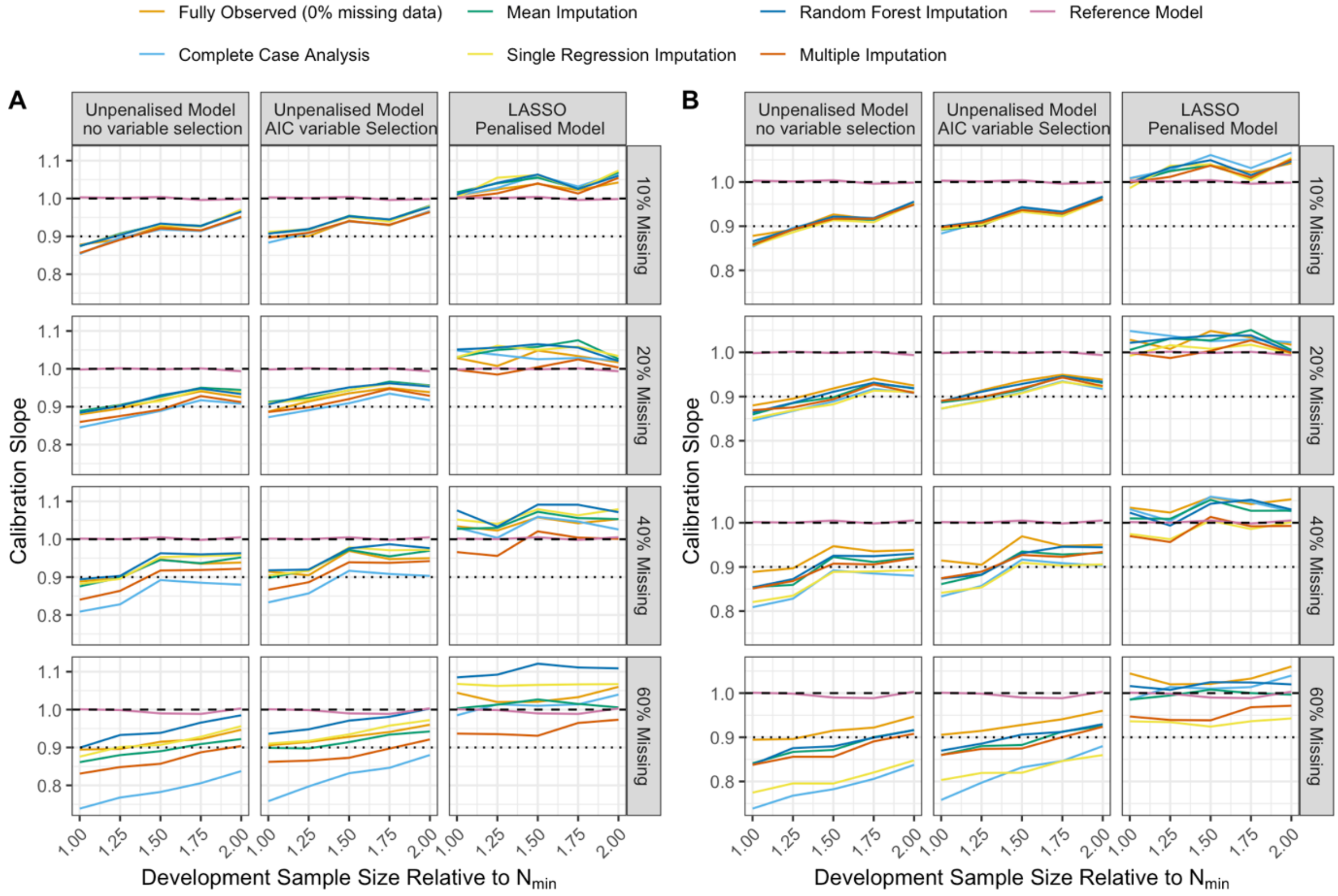

**Supplementary Figure 3:** Median calibration slope for each model across different sample sizes relative to the closed-form minimum sample size calculation, and different levels of missing data. Panel (A) shows the results estimated within the fully observed target population and Panel (B) shows the results estimated within the imputed version of the target population (using the imputation procedure corresponding to each CPM). Results are for cases where data were missing not at random (MNAR), for $\rho = 0.5$ and where only five of the predictor variables are associated with the outcome. The dotted horizonal line shows that targeted value of 0.9, and the dashed horizontal line shows a perfect calibration slope of 1.

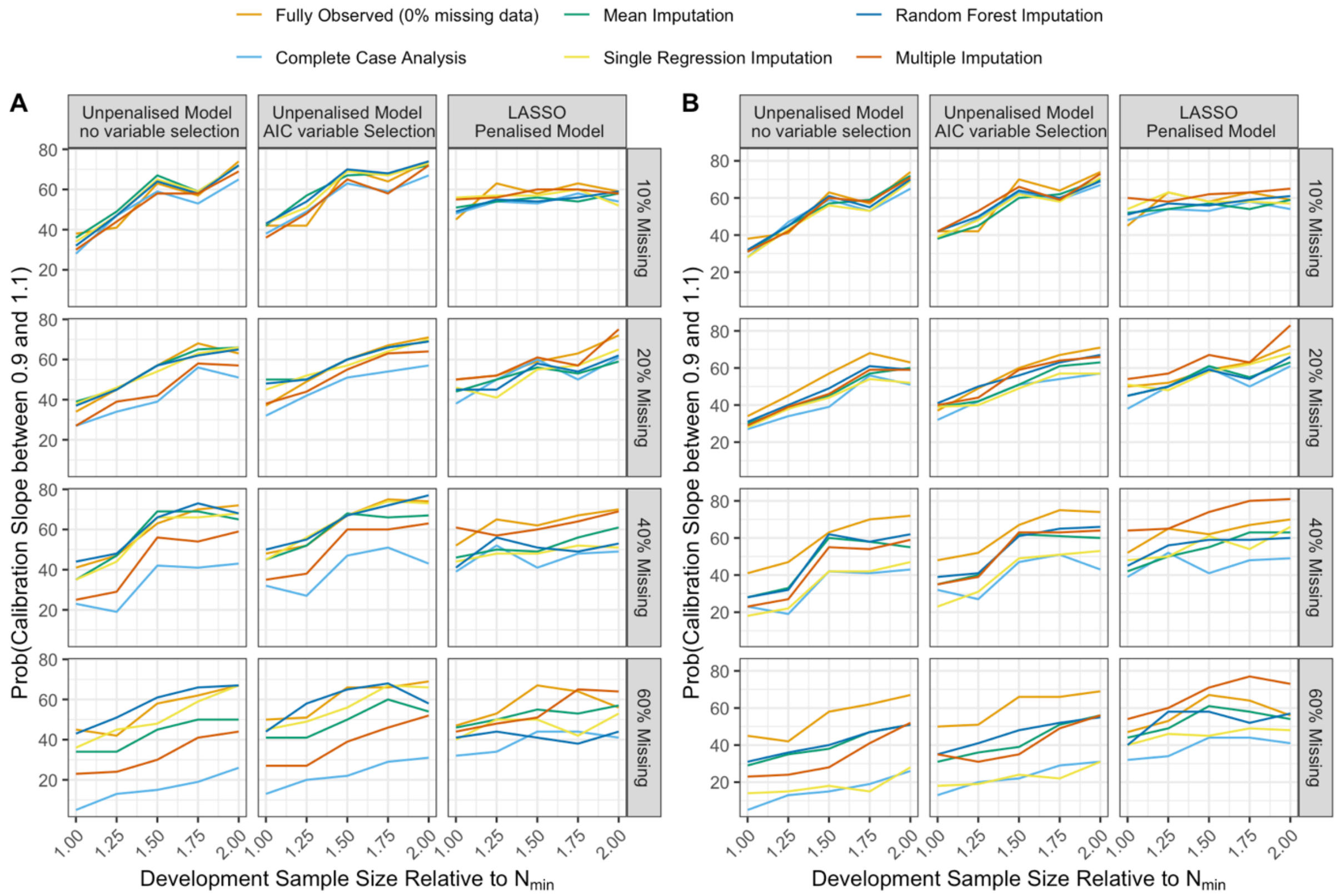

**Supplementary Figure 4:** Posterior probability of the calibration slope being between 0.9 and 1.1 for each model across different sample sizes relative to the closed-form minimum sample size calculation, and different levels of missing data. Panel (A) shows the results estimated within the fully observed target population and Panel (B) shows the results estimated within the imputed version of the target population (using the imputation procedure corresponding to each CPM). Results are for cases where data were missing not at random (MNAR), for $\rho = 0.5$ and where only five of the predictor variables are associated with the outcome.

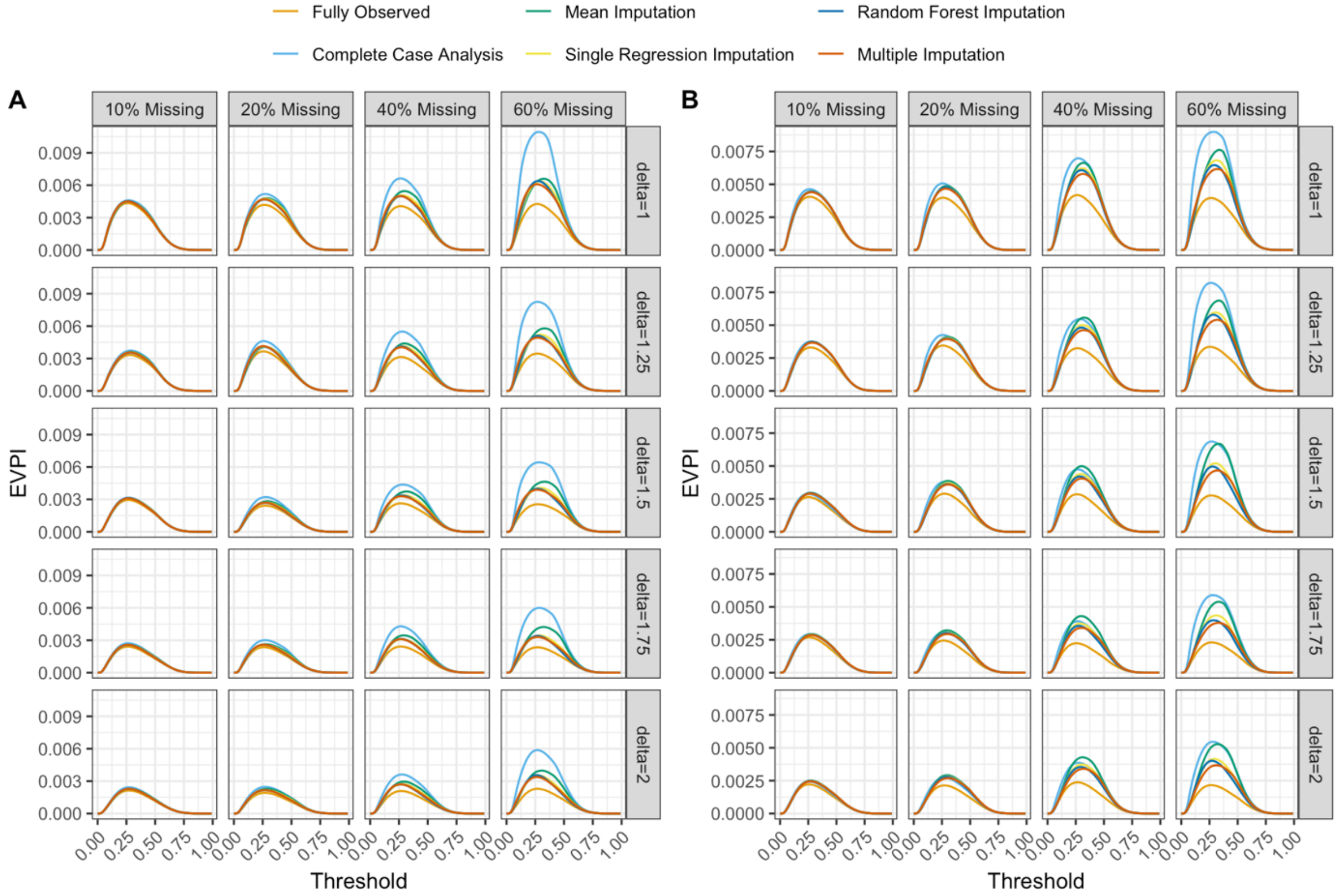

**Supplementary Figure 5:** Expected Value of Perfect Information (EVPI) for each model across different sample sizes relative to the closed-form minimum sample size calculation (rows of each panel; delta), and different levels of missing data. Panel (A) shows the results when data were MAR and Panel (B) shows the results when data were MNAR. Results are for cases where $\rho = 0.5$, where the model was fitted using unpenalised likelihood, and where only five of the predictor variables are associated with the outcome.

## 8 Supplementary Practical 1 Example Results

This section contains supplementary information and results from the practical example one that is described in the main paper.

**Supplementary Table 1:** Summary of the coefficients used for the reference model in practical examples 1 and 2, estimated based on the data in Martin et al. [1].

| **Variable** | **Estimate** |
|---|---|
| (Intercept) | 1.973 |
| Age | |
| <30 | REF |
| 30-40 | 0.228 |
| 40-50 | 0.356 |
| 50-60 | 0.571 |
| 60-70 | 0.767 |
| 70-80 | 0.996 |
| >80 | 1.297 |
| Gender (Female vs. Male) | 0.100 |
| Admission Type (Emergency vs. Elective) | 1.202 |
| Ethnicity | |
| White | REF |
| Black | -0.289 |
| Other | -0.005 |
| Bicarbonate | -0.102 |
| Creatinine | -0.124 |
| Chloride | -0.042 |
| Hemoglobin | 0.018 |
| Platelet | -0.001 |

| Potassium | -0.046 |
|---|---|
| Partial Thromboplastin Time | 0.007 |
| International Normalised Ratio | 0.209 |
| Prothrombin time | -0.007 |
| Blood urea nitrogen | 0.016 |
| White blood cell count | 0.019 |

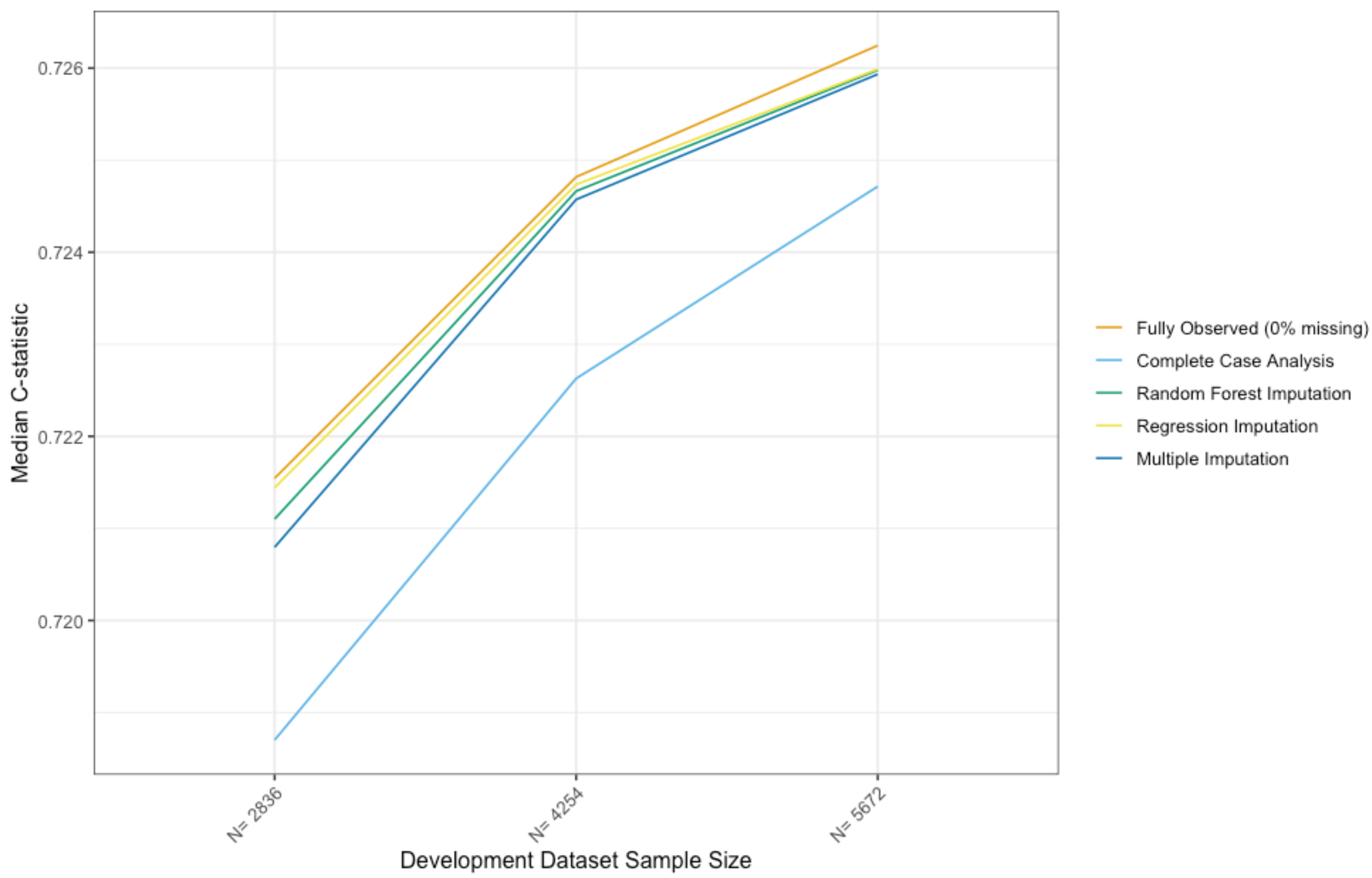

**Supplementary Figure 6:** Median C-statistic for each model fitted within practical example 1.

# 9 Practical Example 2 Methods and Results

In this second practical example, we consider a situation where a researcher has no access to individual participant data (IPD) to help inform the assumptions within the proposed sample size procedure for the design of a new study aiming to develop a prediction model for mortality on intensive care unit (**Figure 1 of main paper**). In such situations, we recommend using external literature, expert opinion, or synthetic reference datasets (where available). For this illustration, we assume only information obtainable from published literature is available to inform the assumptions, and show how the resulting sample size compared with that obtained within example 1.

**Step 1.1:** In the absence of IPD to help inform the joint distributions of predictor variables, we extracted the univariate summaries (i.e., categorical variable frequencies, and mean and variance of continuous variables) of the predictor variables from Martin et al. [1]. There was no published information on the distributions of each variable or on the correlation between variables. Therefore, we here make a simplifying assumption that all continuous variables followed (independent) normal distributions. The reference (true) model was again taken to be that previously reported [1] (**Supplementary Table 1**).

**Step 1.2:** The proportion of overall missing data was assumed to be 20%, based on summary information reported across MIMIC data studies. Patterns of missing data were unknown, so we assumed all combinations of missingness pattern across all variables. In the absence of other information, we conservatively assumed data were MNAR.

**Step 2.1a:** Using the univariate summaries from step 1.1, a large target population was simulated where categorical variables were generated according to the proportions of each category, and continuous variables were generated from a series of independent normal distributions using the mean and variance of each variable. Outcomes were generated from a binomial distribution using the 'true' probabilities from the model in step 1.1.

**Step 2.1b:** As with example 1, we assume we are targeting a CPM where all data will be required upon deployment, so this step is skipped.

**Step 2.2a:** A separate development dataset was generated under the same process as described in step 2.1a above, again using the closed-form procedure of Riley et al. [2] as the initial value of $N_{dev}$ (as described in example 1 above).

**Step 2.2b:** To induce missingness into the dataset generated from step 2.2a, we used fully conditional specification in multivariate imputation [3], implemented using the *ampute* function within the *mice* R package [4], using the assumptions described in step 1.2.

**Steps 3.1 – 5.1:** These steps proceed as described in example 1 above.

**Steps 5.2 & 5.3**: At the initial sample size of 2836, we found similar median calibration slopes to those in example 1 (**Supplementary Figure 7**). The median calibration slope was 0.83, 0.87, 0.87 and 0.86 for the model developed under complete case analysis, random forest imputation, single regression imputation and multiple imputation, respectively. Again, the assurance probabilities were generally lower for all models. The median C-statistic of all models was also lower than in example 1 at 0.71 (**Supplementary Figure 8**).

**Step 5.4**: As with example 1, increasing the sample size to 1.5 times that from closed-form solutions, resulted in a median calibration slope of all models (except complete case analysis) being above 0.9, with assurance probabilities above 50% (**Supplementary Figure 7**). It is reassuring that we arrive at similar conclusions across the two examples, despite the differences in underlying assumptions. In practice, we would recommend that research perform sensitivity analyses around assumptions to test how different the sample size results become.

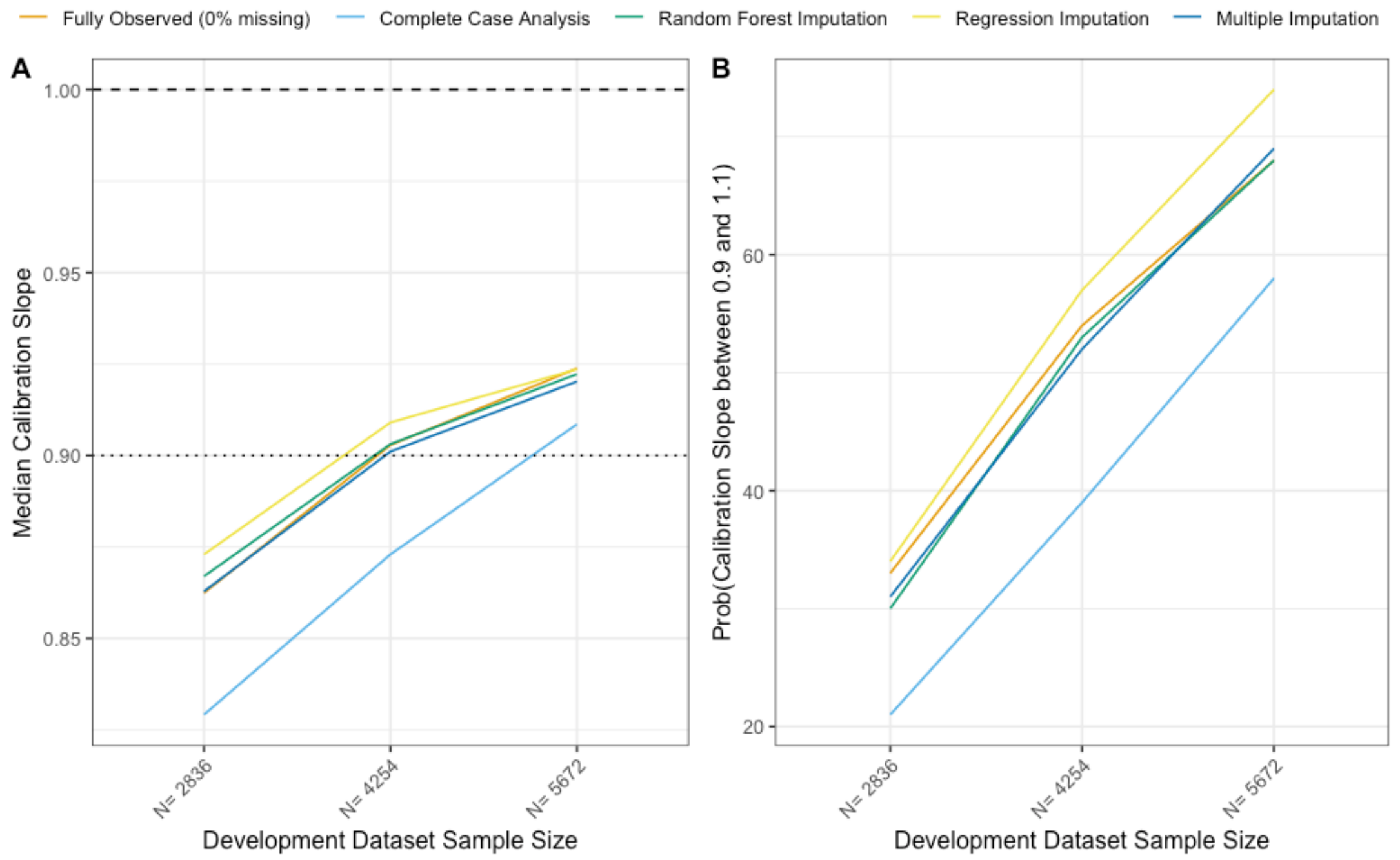

**Supplementary Figure 7:** Median calibration slope (Panel A) and assurance probability (Panel B) for each model across different sample sizes for practical example 2.

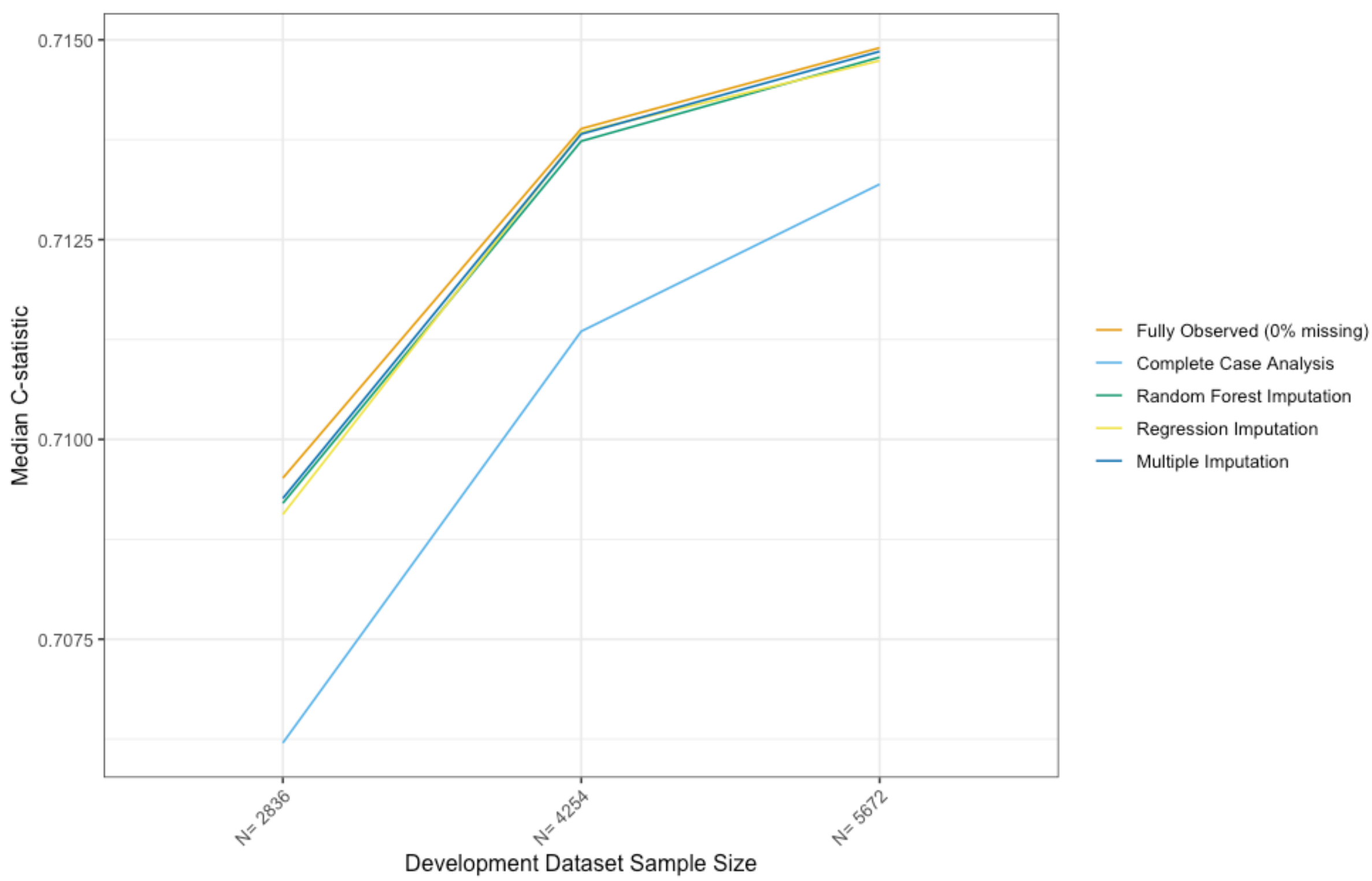

**Supplementary Figure 8:** Median C-statistic for each model fitted within practical example 2.

## 10 Supplementary References